%% file: fluent.tex
\documentclass[conference]{IEEEtran}

\IEEEoverridecommandlockouts
 
\usepackage{algorithmic}
\usepackage{graphicx}
\usepackage{textcomp}
\usepackage{xcolor}
\usepackage{booktabs} 
\usepackage{listings}
\usepackage{multirow}
\usepackage{soul}
\input{preamble} 

\definecolor{source}{gray}{0.95}
\usepackage[T1]{fontenc} 

\usepackage{caption}
\usepackage{subcaption}

\usepackage{pgfplots} 
\usetikzlibrary{patterns}

\newcommand{\boxit}[1]{\vspace{0.3cm}
\noindent
\fbox{
\begin{minipage}{8.3cm}
\emph{#1} 
\end{minipage}
}
}

\lstdefinelanguage{Java}{
  tabsize=4
}[keywords,comments,strings]

\definecolor{source}{gray}{0.95}
\definecolor{highlight}{gray}{0.9}

\usepackage{courier} 
\usepackage{multicol} 
\usepackage{caption}

\lstdefinelanguage{JavaScript}{
  keywords={let, const, typeof, new, true, false, catch, function, return, null, catch, switch, var, if, in, while, do, else, case, break},
  keywordstyle=\color{blue}\bfseries
 }
 
 \lstdefinelanguage{CryRule}{
   sensitive = false,
  keywords={ALGORITHM, in, length, iv, symmetrickey },
  keywordstyle=\color{blue}\bfseries,
  keywords=[2]{Cipher, Hash},
  keywordstyle=[2]\color{black}\bfseries
 }

\lstnewenvironment{codesnippet}{%
	\lstset{%
		frame=single,
		framerule=0pt,
		mathescape=false
	}
}{}

\makeatletter
\lst@Key{countblanklines}{true}[t]%
    {\lstKV@SetIf{#1}\lst@ifcountblanklines}

\lst@AddToHook{OnEmptyLine}{%
    \lst@ifnumberblanklines\else%
       \lst@ifcountblanklines\else%
         \advance\c@lstnumber-\@ne\relax%
       \fi%
    \fi}
\makeatother

\newcommand{\GH}{GitHub\xspace}

\newcommand{\SO}{Stack Overflow\xspace}

\newcommand{\NJ}{Node.js\xspace}
\newcommand{\FC}{FluentCrypto\xspace}

\begin{document}

\title{FluentCrypto: Cryptography in Easy Mode}

\author{
\IEEEauthorblockN{Simon Kafader}
\IEEEauthorblockA{University of Bern}
Bern, Switzerland\\
simon.kafader@inf.unibe.ch
\and
\IEEEauthorblockN{Mohammad Ghafari}
\IEEEauthorblockA{University of Auckland\\
Auckland, New Zealand\\
m.ghafari@auckland.ac.nz}
}

\maketitle

\begin{abstract}

Research has shown that cryptography concepts are hard to understand for developers, and secure use of cryptography APIs is challenging for mainstream developers. 
We have developed a fluent API named \FC to ease the secure and correct adoption of cryptography in the \NJ JavaScript runtime environment. 
It provides a task-based solution i.e., it hides the low-level complexities that involve using the native \NJ cryptography API, and it relies on the rules that crypto experts specify to determine a secure configuration of the API.
We conducted an initial study and found that \FC is hard to misuse even for developers who lack cryptography knowledge, and compared to the standard \NJ crypto API, it is easier to use for developers and helps them to develop secure solutions in a shorter time.

\end{abstract}


\section{Introduction}
\label{cha:introduction}

The estimated cost of a data breach in the United States is on average USD 8.64 million~\cite{ibm}.
Most software systems rely on cryptography to protect sensitive data, and developers, who often are not security experts, should build these systems.

Despite huge efforts in the field of usable security in general and cryptography in particular, cryptography APIs (in short, crypto APIs) are still difficult to use for mainstream developers~\cite{Patnaik2020}, resulting in crypto API misuses and exposing otherwise sensitive data in practice.
A recent study of 2 324 open-source Java projects on GitHub showed that 72\% of the projects suffer from at least one crypto misuse, and that the frequency of using crypto APIs does not change developer performance in this domain~\cite{hazhirpasand2019impact}.
There are a number of reasons for poor adoption of cryptography.
Foremost, security caveats are rare in the documentation of crypto APIs \cite{Hazhirpasand2020}.
Developers may instead consult online information sources such as the Stack Overflow website to learn about cryptography and adopt online code examples in their programs.
Unfortunately, such sources are not reliable either.
For instance, 
a study of 217 818 Stack Overflow posts revealed that 31\% of them suffer from API misuses, which has this possibility to leak resources \cite{zhang2018code}.
There exist several tools to detect security issues in general~\cite{Ghafari2017, Gadient2019}, and crypto-related misuses in particular~\cite{kruger2017cognicrypt, Rahaman2019}.
However, research has shown that these tools are hard to use for mainstream developers~\cite{Tymchuk2018, Smith2020}, and security analysis tools perform differently in practice~\cite{Corrodi2018, Ranganath2020}.
Finally, there are dedicated platforms to explore real-world crypto code examples~\cite{hazhirpasand2020cryptoexplorer}, but it is still a developer's responsibility to adapt these examples in her program, which may be an error-prone task.
%
%

Unlike previous work that mostly supported developers to detect crypto misuses, in this paper, we aim to ease the secure adoption of crypto APIs in the first place \ie during programming.
We introduce \FC, a wrapper around the \NJ crypto API, that bridges the gap between cryptography experts and mainstream developers.
It relaxes developers from crypto concerns such as the order of API calls, the choice of a secure algorithm, the right way to generate a secret key, \etc
\FC utilizes the fluent interface design principle to hide such low-level crypto-related complexities \ie developers only state ``what'' they need without being concerned about ``how'' to do it.
It relies on the constraints that crypto experts specified to determine a correct and secure configuration of the API, and when something goes wrong, it provides developers with guiding error messages to resolve the issue.
In order to understand whether or not \FC eases secure adoption of cryptography in practice, we asked two research questions:

\begin{itemize}

\item
\textbf{RQ$_1$}: Does adopting \FC increase developer efficiency when working on a cryptographic task?

\item
\textbf{RQ$_2$}: Is \FC effective in developing secure cryptographic solutions?

\end{itemize}

To answer these questions, we evaluated \FC by conducting an initial study with eight participants from the software industry.
Each participant had to complete three tasks in cryptography with the help of \NJ cryptography API and \FC.
The results revealed that \FC is greatly helpful in terms of decreasing the security risks, effort and time required to solve a cryptography task.
In particular, more participants were able to complete the given tasks with \FC compared to the same tasks with the \NJ crypto API.
Importantly, less experienced participants finished the tasks much faster when they used \FC than when they relied on the \NJ crypto API.
Notably, all participants provided secure solutions when they used \FC.


The rest of this paper is organized as follows.
We discuss related work in \autoref{sec:relatedWork}.
In \autoref{sec:motivation}, we demonstrate an example concerning why adopting cryptography can be problematic for an inexperienced developer. 
In \autoref{sec:fluent}, we present \FC and a domain-specific language that we developed for crypto experts in order to specify constraints on crypto objects.
In \autoref{sec:validationSetup}, we describe the setup of our evaluation, and we present the evaluation result in \autoref{sec:validationResult}.
We discuss the threats to validity of this work in \autoref{sec:threats} and conclude the paper in \autoref{sec:conclusion}.

\section{Related Work}
\label{sec:relatedWork}

We discuss related work from two perspectives. 
Firstly, we discuss an excerpt of literature that investigated why developers struggle with using crypto APIs.
Secondly, we discuss previous work that aimed to help developers in circumventing crypto misuses.
Nevertheless, to the best of our knowledge, there is no such research on \NJ cryptography.

\subsection{API design}

Nadi \etal conducted a survey with 11 developers who posted crypto-related questions on the Stack Overflow website, and 37 developers who had experience with Java's cryptography APIs \cite{nadi2016jumping}.
They realized that developers are able to understand crypto concepts but they fail to adopt them in practice.
They concluded that the concepts behind crypto APIs are low-level and developers welcome task-based solutions for working with crypto APIs.

Green and Smith found that many misuses of cryptographic libraries originate in the developer having trouble understanding the API \cite{green2016developers}. 
They criticized that while it has become accepted that systems should be user-friendly for the end user, a different attitude where the end-user is expected to be an expert prevails amongst cryptographic libraries. 
They then proposed a set of characteristics that should lead to more secure cryptographic APIs, including: 
the API should be easy to use, even without documentation; 
incorrect use should lead to visible errors;
defaults should be safe and never ambiguous;
and code that uses the API should be easy to read and update.

Das \etal  selected the most popular cryptographic libraries from C, C++, Java, Python and Go and examined them for properties that cause cryptographic misuse by developers \cite{das2014iv}. 
They examined the libraries based on a set of common potential issues such as initialization vector reuse, library defaults or incomplete features.
Their work highlighted the disconnect between the actual user of such a library and the user for whom it is designed. 

\subsection{Developer support}

Several tools have been developed to support developers in using cryptography. 
For example, Krüger \etal developed the CogniCrypt Eclipse plug-in that generates code snippets for cryptography tasks~\cite{Kruger2020}. 
Nguyen et al developed FixDroid which helps developers in ﬁxing crypto-related issues in the AndroidStudio IDE~\cite{Nguyen:2017}.
Hazhirpasand \etal developed an interactive web platform named CryptoExplorer that provides developers with real-world examples, specifically 3263 secure and 5897 insecure uses of Java Cryptography Architecture~\cite{hazhirpasand2020cryptoexplorer}.
Singleton \etal developed CryptoTutor, an educational tool that flags common cryptographic misuses and suggests possible repairs~\cite{Singleton2020}. 
The authors discussed how such a tool can be integrated into programming courses to improve developer knowledge in the cryptography domain.

Several tools also exist to detect crypto misuses.
For instance, 
Rahaman \etal developed CryptoGuard to identify crypto issues statically~\cite{Rahaman2019}.
Piccolboni \etal proposed CryLogger, an open-source tool to detect crypto misuses  dynamically~\cite{Piccolboni2021}.
Gorski surveyed 25 professional software developers to identify what kind of feedback is helpful to them in avoiding crypto misuses during programming~\cite{Gorski2020}.
They found that participants appreciate a clear warning message (including title, content, and code location) that is tailored to the context.

\section{Background and Motivation}
\label{sec:motivation}

\NJ is a popular open-source, cross-platform, back-end JavaScript runtime environment to build scalable network applications.\footnote{https://nodejs.org}
It provides a built-in library called ``crypto'' which developers can use to perform cryptographic operations on data.
According to the latest survey conducted by the Stack Overflow website in 2020,\footnote{\href{https://insights.stackoverflow.com/survey/2020\#technology-other-frameworks-libraries-and-tools}{Stack Overflow Developer Survey 2020}}
for the second year in a row, \NJ is a worldwide leader among frameworks.
However, to the best of our knowledge, there exists no work in academia to address crypto misuses in this environment.

\begin{lstlisting}[
countblanklines=false, 
float=*,
xleftmargin=2em, frame=single, framexleftmargin=15pt, belowcaptionskip=8pt,
caption=An example of insecure symmetric encryption, label=lst:exampleEncryption]

const crypto = require('crypto');
 
const algorithm = 'des';
const key = crypto.scryptSync('password', 'salt', 8);
const iv = Buffer.alloc(8,0);
 
const cipher = crypto.createCipheriv(algorithm, key, iv);
 
const ciphertext = cipher.update('some clear text', 'utf8', 'hex');

cipher.final();

\end{lstlisting}

Consider \autoref{lst:exampleEncryption} that shows a code example to encrypt data with a private key.
We access the cryptography module at line 1.
We specify the encryption algorithm \ie ``des'', derive the encryption key using the \texttt{Scrypt} function, and define the initialization vector (iv) at line 2-4, respectively.
Instances of the \texttt{Cipher} class are used to encrypt data, and we create and initialize such an instance at line 5.
The \texttt{cipher.update()} method performs the actual encryption.
It takes three arguments \ie the data, the input encoding, and the output encoding.
 We call this method at line 6 to encrypt a piece of ``utf-8''-encoded text and store the result in the hex format.
In the end, at line 7, we call the \texttt{cipher.final()} method to indicate that the \texttt{Cipher} object is no longer needed and the encryption process is finalized.

There are a number of caveats that developers should take into account to ensure that this code and its counterpart for decryption work correctly.
For instance, 
the default string encoding used by the crypto module is utf-8.
We also need a key of a certain length, depending on which cipher algorithm we choose for encryption.
In this example, the iv and the key must have the same length \ie number of bytes.
%
If the encoding of the input data is given to the \texttt{cipher.update()} method, the data argument is a string in the specified encoding. 
Otherwise, data must be a Buffer.
Likewise, if the encoding of the output is specified, a string in the specified encoding is returned. Otherwise, a Buffer is returned.
%

Even if we assume that a novice developer will figure out how to write the code in \autoref{lst:exampleEncryption} that is functioning, it suffers from a number of security issues.
We used \texttt{Scrypt}, a password-based key derivation function, which requires three arguments namely password, salt and the key length to construct the secret key.
A salt is combined with the password to avoid producing an identical key for the same password, which substantially lessens the impact of pre-computed hash attacks especially against commonly used passwords\footnote{\href{https://nvlpubs.nist.gov/nistpubs/Legacy/SP/nistspecialpublication800-132.pdf}{NIST recommendation for password-based key derivation}}.
Therefore, salt should be as unique as possible and it is recommended that salt is random and at least 16 bytes long.
However, in the example, the salt is a fixed value and does not help in deriving distinct keys.
%
%
The next security issue concerns the IV whose purpose is to ensure that an identical message encrypted with the same key leads to different encrypted outputs.
This requires the IV to be unpredictable and unique; and ideally, cryptographically random so that an attacker cannot predict ahead of time what a given IV will be.
Nevertheless, in the example, the IV is always a zero-filled buffer.
%
Finally, DES is no longer a secure algorithm due to its comparatively short key size.

We investigated obstacles as well as security risks that are associated with hashing, symmetric and asymmetric encryption in \NJ.
We decided to focus on these cryptographic tasks as research as well as our experience show that they are commonly used to secure software systems~\cite{Hazhirpasand2020}.
It is notable that a developer must have knowledge about, for instance, the choice of algorithm, the correct key and IV generation process, and input and output encoding.
Even worse, error messages are sometimes cryptic.
For example, 
by default, \NJ crypto API assumes that the input encoding must be a buffer, and the failure in setting the right character encoding during decryption yields the  ``\emph{error:0606506D:digital envelope routines:EVP\_DecryptFinal\_ex:wrong final block length}'' error, which does not guide developers per se toward resolving the issue.

It would have been much easier if developers could have just told the API that, for example, they want to encrypt data and then decrypt it later, without necessarily having to search for a (secure) algorithm; understanding how long the key and the IV have to be, and how to generate them securely; and with more useful error messages when something goes wrong.

\section{FluentCrypto}
\label{sec:fluent}

We developed \FC, a wrapper around the \NJ crypto API, with the aim of easing the secure adoption of crypto APIs for mainstream developers.
\FC bridges the gap between cryptography experts and developers.
It relaxes developers from crypto concerns such as the order of API calls, the choice of a secure algorithm, the right way to generate a key, etc, and when something goes wrong, it provides developers with guiding error messages to resolve the issue.

\begin{figure*}
  \includegraphics[width=\linewidth]{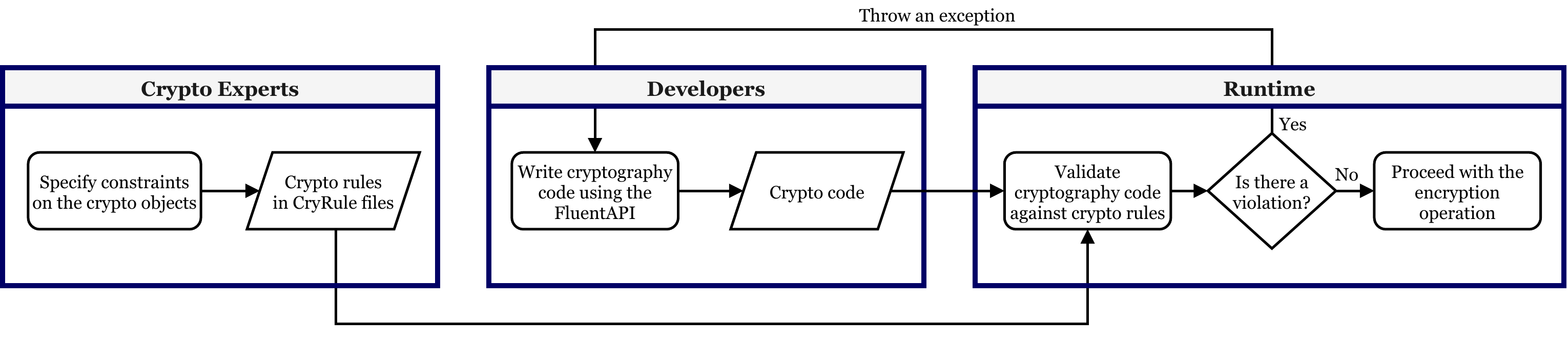}
  \caption{FluentCrypto work flow}
  \label{img:flow}
\end{figure*}

The work flow of \FC is shown in \autoref{img:flow}.
First, crypto experts should specify the secure configuration of crypto objects (\ie security constraints) in a domain specific language, named \emph{CryRule}, that we designed for this purpose.
Then, developers write crypto code using the \FC that enables defining a crypto task and providing necessary configurations in a method-chaining style.
%
%
\FC is built to be hard to misuse.
It hides the complexities behind the scene \ie developers only specify ``what'' they need without concerning about ``how'' to do it.
Yet, it allows expert developers to provide their own crypto objects, if needed.
Finally, \ie at runtime,
\FC initializes each crypto object with secure defaults that experts specified, and it validates each configuration provided by a developer against the latest security constraints to determine a correct use of cryptography.
We employed a dynamic approach mainly due to the use of JavaScript, but it is worthwhile to investigate the possibility of adopting statical techniques to provide developers with JIT feedback especially in other settings.

\FC currently supports hashing, symmetric as well as asymmetric encryption. 
In the rest of this section, we only discuss \emph{symmetric encryption}, which is known to be one of the most commonly used types of cryptography among developers~\cite{Hazhirpasand2020}.
\FC is publicly available on GitHub.\footnote{https://github.com/Smoenybfan/FluentCrypto}.

\subsection{CryRule}


We developed a domain-specific language that crypto experts can use to specify secure configurations of cryptography objects independent from the actual cryptography code.
We named this language ``CryRule'' and refer to its files (.cryrule) as ``rule files''.
We built CryRule using the chevrotain JavaScript library, an open-source parser building toolkit.\footnote{https://github.com/SAP/chevrotain}
We also developed a parser that compiles CryRule files into plain JavaScript objects that we use to validate the crypto-related configurations at runtime and determine the secure use of crypto APIs.

There are many ways that developers can use a crypto API but only a few correspond to correct and secure usages.
We therefore tailored CryRule to use \emph{white listing} \ie the rules indicate secure configurations whereas all deviations from such rules are assumed to be insecure.
Each rule specifies constraints on the values of a parameter that a crypto API can hold.
CryRule is not whitespace sensitive; neither line breaks nor indentations change the meaning.


%
A rule file in CryRule includes one or multiple sections.
Each section describes the constraints on a certain cryptographic task, and starts off by specifying the name of individual classes, such as \texttt{Cipher} or \texttt{Hash}, that the crypto constraints apply to.

There are a few keywords in CryRule to define a rule.
For instance, 
\textsc{\textbf{algorithm}} defines the algorithm(s) that are allowed for a task.
This keyword is followed either by a single algorithm name (\eg \autoref{lst:algorithmCipher}) or by the \textsc{\textbf{in}} keyword and a list of algorithms inside the square brackets (\eg \autoref{lst:multipleAlgorithm}).

\noindent
\begin{minipage}{\linewidth}
\begin{lstlisting}[language=CryRule, numbers=none, caption=Single algorithm whitelisting, label=lst:algorithmCipher]
Cipher
	ALGORITHM aes-128-cbc
\end{lstlisting}
\end{minipage}

\noindent
\begin{minipage}{\linewidth}
\begin{lstlisting}[language=CryRule, numbers=none, caption=Multiple algorithm whitelisting, label=lst:multipleAlgorithm]
Cipher
	ALGORITHM IN [aes-128-cbc, aes-128-gcm, aes-192-cbc, aes-192-gcm, aes-256-cbc, aes-256-gcm]
\end{lstlisting}
\end{minipage}

\textsc{\textbf{length}} defines a constraint on an object that has a length, \eg a symmetric key or an IV. 
This keyword is followed by either a number or the keyword \textsc{in} and an array of numbers, which are the lengths that the object is allowed to accept. 
The number(s) can be followed by an \textsc{\textbf{if}} statement that declares the condition under which the specified constraint (in this case, the length) is valid. 
For example, \autoref{lst:iv} specifies the constraints on the \textsc{\textbf{iv}}.
In particular, it states that the IV should be 16 bytes if the cipher algorithm is ``aes-128-cbc'', ``aes-192-cbc'', or ``aes-256-cbc''; and it must be 12 bytes if the cipher uses the ``aes-128-gcm'', ``aes-192-gcm'', or ``aes-256-gcm'' algorithm.

\noindent
\begin{minipage}{\linewidth}
\begin{lstlisting}[language=CryRule, numbers=none, caption=Constraints on the length of initialization vector, label=lst:iv]
Cipher
	IV
	LENGTH 16 IF ALGORITHM IN [aes-128-cbc, aes-192-cbc, aes-256-cbc]
	LENGTH 12 IF ALGORITHM IN [aes-128-gcm, aes-192-gcm, aes-256-gcm]
\end{lstlisting}
\end{minipage}

It is possible to specify constraints for a symmetric key using the \textsc{\textbf{symmetrickey}} keyword.
It consists of multiple \textsc{length} constraints on the length of the key w.r.t. the algorithm used in the key generation process; a \textsc{\textbf{saltlength \textgreater=}} constraint to set the minimum length for the salt; and finally, an \textsc{\textbf{iterations \textgreater=}} constraint to set the minimum times that the key generation algorithm must repeat to generate a symmetric key.\footnote{The order of constraints on a \textsc{symmetrickey} is important.}
The CryRule snippet presented in \autoref{lst:symKeyCon} shows constraints on a symmetric key.
The length of the key is bound to 16 bytes if the algorithm for the cipher task is either ``aes-128-cbc'' or ``aes-128-gcm''; it is 24 if the cipher uses either ``aes-192-cbc'' or ``aes-192-gcm'' algorithm; and it must be 32 in case of ``aes-256-cbc'' or ``aes-256-gcm'' algorithm.  
The minimum number of iterations is set to be 10 000, and the salt is restricted to a length of at least 20 bytes.

\noindent
\begin{minipage}{\linewidth}
\begin{lstlisting}[language=CryRule, numbers=none, caption=Symmetric key constraints, label=lst:symKeyCon]
Cipher
	SymmetricKey
	LENGTH 16 IF ALGORITHM IN [aes-128-cbc, aes-128-gcm]
	LENGTH 24 IF ALGORITHM IN [aes-192-cbc, aes-192-gcm]
	LENGTH 32 IF ALGORITHM IN [aes-256-cbc, aes-256-gcm]
        
	ITERATIONS >= 10000
  
	SALTLENGTH >= 20
    
\end{lstlisting}
\end{minipage}

It is worth noting that 
we derived the current set of rules based on our expertise and by consulting literature~\cite{Kruger2020}, however, experts can always update these constraints according to the latest standards and security advice.

\subsection{The fluent interface}

We developed, FluentCrypto, an API that enables easy and secure adoption of cryptography.
It utilizes the fluent interface design principle to achieve method chaining,~\ie invoking multiple methods successively, so that it reads very similar to a natural-language text~\cite{FAPI}.
%
The order of method calls does not matter, and methods are named after the tasks they solve.
\FC ensures a correct and secure adoption of cryptography in two ways.
%
Firstly,
developers only specify ``what'' they need without concerning about ``how'' to do it.
Consequently, 
it clears implementation complexities such as 
calling the wrong methods on the API objects, calling methods in an incorrect order, or missing to call the methods entirely.
For instance, 
\FC uses a secure source of randomness and generates a new IV for every encryption.
Secondly, 
\FC relies on the constraints that crypto experts specify to determine a correct and secure configuration of the API such as the choice of algorithm and the length of an IV.
Specifically, 
in the initialization of the encryption process, it parses the latest crypto constraints that are specified in the CryRule files and passes these rules to the related objects.
When a crypto API function is called, all the associated rules are validated against the current configuration and an exception is thrown in case of a violation.

\autoref{lst:basicEncryption} presents a very basic example of using \FC to perform encryption. 
In the first line, we access the \FC library.
We tell the API that we want to perform an encryption task in the third line, and provide it with the plain input data in the fourth line.
Finally, we call the \texttt{run} method to perform the actual encryption.

\begin{lstlisting}[language=JavaScript, 
%numbers=none, 
countblanklines=false,
caption=Encryption with \FC, label=lst:basicEncryption]
const FluentCrypto = require('./FluentCrypto');

const encrypted = FluentCrypto
		  .encryption()
		  .data('some plain text')
		  .run();

\end{lstlisting}

Even in the simplest encryption scenario, \eg to encrypt data with a secret key, a developer should provide the encryption algorithm and the key.
In order to facilitate the adoption of cryptography, \FC uses secure defaults for any input that is missing.
A secure default is the first value that is associated with each constraint in the CryRule files.
For instance, considering the code example in \autoref{lst:basicEncryption} and the constraints in \autoref{lst:multipleAlgorithm}\autoref{lst:iv}\autoref{lst:symKeyCon}, \FC would use ``aes-128-cbc'' algorithm; securely generate the right sizes of a key and an IV (\ie 16 bytes); identify and set the right encoding \ie utf-8.
\FC uses \texttt{pbkdf2} as its default key derivation function.

\newpage
\FC provides a set of functions that allow developers to specify their own configurations and set up crypto objects manually, similarly to configuring the native API.
A number of these functions are listed in~\autoref{tbl:functions}.\footnote{Please consult~\href{https://github.com/Smoenybfan/FluentCrypto}{the \FC project on GitHub}
to find a complete list of these functions.} 
For instance, the \texttt{withCipher(}`aes-192-cbc'\texttt{)} method instructs the API to use the AES-192 bit Cipher algorithm in Cipher Block Chaining (CBC) mode.
A developer can set a key using the \texttt{withSymmetricKey()} method.
If the given key features are not according to the chosen cipher algorithm, \FC will throw an error.
However, it should be noted that when a developer inputs her own crypto object, such as a secret key, 
\FC cannot any longer determine whether the key is secure.
Precisely, \FC runs the checks on the provided key \ie the length constraint, but it cannot examine if the key was derived in a secure way.

\begin{table*}[t]
\centering
\caption{\FC configuration functions for symmetric encryption}
\label{tbl:functions}
\begin{tabular}{|l|p{10cm}|}
\hline
\rowcolor[HTML]{BBE0D7} 
\textbf{Name} & \textbf{Description} \\ \hline

	\texttt{withCipher	}&Instructs the API to use symmetric encryption with a cipher and initializes the defaults.
	If the user provides a specific algorithm, it uses that algorithm instead of its default.\\
	\hline
	
	\texttt{withCipherfromPassword}&Instructs the API to use symmetric encryption with a cipher, initializes the defaults and generates a key from the provided password.\\
	\hline
	
	\texttt{withCipherFromSymmetricKey}&Instructs the API to use symmetric encryption with a cipher, uses the user's provided key and initializes the defaults for that key.\\
	\hline
	
       \texttt{setKey}&Instructs the API to use the provided key for symmetric encryption, and overwrites the default.\\
       \hline
       
       \texttt{setIV}&Instructs the API to use the provided initialization vector for symmetric encryption, and overwrites the default.\\
       \hline
       
       \texttt{setKeyGenerationPassword}&Sets and overwrites the password that the default symmetric key is generated from to the provided password. Triggers the key generation and overwrites the set or default key.\\
       \hline
       
       \texttt{setKeyGenerationSalt}&Sets and overwrites the salt the default symmetric key is generated with to the provided salt. Triggers the key generation and overwrites the set or default key.\\
       \hline
       
       \texttt{setKeyGenerationIterations}&Sets and overwrites the amount of iterations that the default symmetric key is generated with to the provided iterations.
       Triggers the key generation and overwrites the set or default key.\\
       \hline
       
       \texttt{setSymmetricKeyGenerationAlgorithm}&Sets and overwrites the digest algorithm with which the default symmetric key is generated to the provided algorithm. Triggers the key generation and overwrites the set or default key.\\
       \hline
       
\end{tabular}
\end{table*}

Finally, in order to perform decryption, developers should simply invoke the \texttt{Decryption} function and provide all necessary inputs as shown in \autoref{lst:basicDecryption}.
It is worth mentioning that \FC provides a set of getter methods, such as \texttt{getKey} and \texttt{getIV}, to access crypto-related objects that are used in an encryption task. These objects are required for a successful decryption operation.

\noindent
\begin{minipage}{\linewidth}
\begin{lstlisting}[language=JavaScript, 
numbers=none, 
caption=Decryption with \FC, label=lst:basicDecryption]
const FluentCrypto = require('./FluentCrypto');

const decrypted = FluentCrypto.decryption()
		.data(cipher text)
		.withCipher(algorithm)
		.key(key)
		.iv(iv)
		.run()
\end{lstlisting}
\end{minipage}


\section{Study setup}
\label{sec:validationSetup}

\begin{figure}[h]
	\centering
  \includegraphics[width=0.4\textwidth]{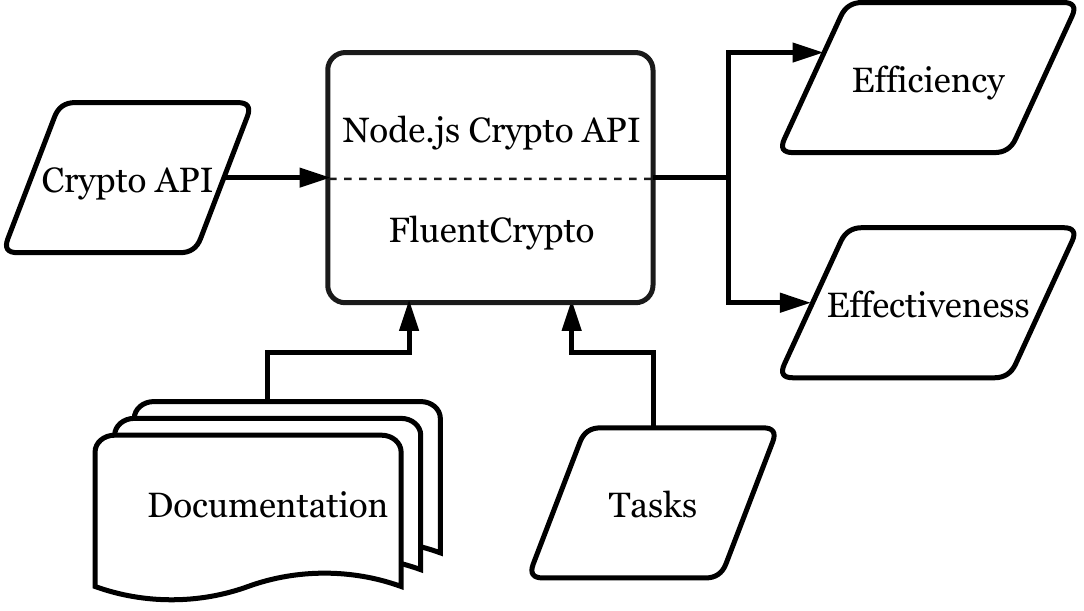}
  \caption{The study setup}
  \label{img:experiment}
\end{figure}

We performed
a study, outlined in \autoref{img:experiment}, 
to evaluate \FC.
We asked participants to complete three basic cryptographic tasks using the \NJ crypto API and \FC.
We designed the tasks based on real-world problems found on the Stack Overflow website.
%
%
We measured the results by asking the participants to record how long they spent on each task, state a difficulty level and explain the impediments they had while working on each task.
We also asked the participants to send us their final code pieces for further investigations.
We manually inspected their implementation to assess the characteristics of their solutions in terms of both completeness,~\ie delivering what a task requests, and correctness,~\ie providing a secure solution.

\subsection{Tasks}

We designed three tasks based on the frequent problems that developers recently asked about hashing, symmetric encryption, and key-pair encryption on the Stack Overflow website. 

\subsubsection{Task 1} 
We searched for the ``hash'' and ``node.js'' tags on Stack Overflow, which resulted in more than 30 active questions within the last week at the time of running this experiment.
We skipped the questions that were not related to problems tied with \NJ crypto API. 
We analyzed the first ten related questions and realized that seven of askers did not know how to use a hash algorithm; one did not have adequate knowledge to understand hashing; and the remaining were related to libraries.
We therefore designed the following symmetric encryption task: 

\boxit{Task 1: hash a string using the standard \NJ crypto library and with the help of the \FC.}

The main goal was to comprehend how difficult it is to perform hashing with each API.

\subsubsection{Task 2} 
We searched for the ``cipher'' and ``node.js'' tags on Stack Overflow, which resulted in 10 relevant active questions within the last week at the time of running this experiment.
Five questions were about lack of knowledge in using the API that yield error or incorrect result, two were about problems with concatenating the results from the calls, two questions were related to the correct length of IV, and one was a high-level question about the cipher.
Consequently, we formulated the following task:

\boxit{
Task 2: first encrypt the string ``Secrets that are sent over an insecure channel''.
Then, update the encrypted data with ``must be encrypted at all costs!''. 
Next, decrypt the encrypted data. 
The encryption and decryption must be performed with a symmetric key. 
The encrypted data must be in hex encoding, and the plain data in utf-8 encoding.
}

We were interested to comprehend how difficult it is to perform encryption and decryption with a symmetric key (cipher) in \NJ crypto API and \FC.

\subsubsection{Task 3} We searched for ``private encrypt'' and the ``node.js'' tags, and the results contained 251 questions.  
We analyzed the first top ten questions that were relevant. Eight such questions were about how to generate a key pair for asymmetric encryption or how to use the key correctly in the code, one question was about how to set a password for a private key, and one was related to the API's difficulty.
Consequently, we designed the task below:

\boxit{
Task 3: 
Encrypt the string ``Secrets that are sent over an insecure channel must be encrypted at all costs!''.
Next, decrypt the encrypted data. 
The encryption must be performed with a private or public key.
The encrypted data must then be decrypted with corresponding public or private key.
}

The aim of this task was to compare how challenging it is to encrypt/decrypt data with a key pair.

\subsection{Participants}
In order to find interested participants, we wrote two blog posts that describe how \FC works.
%
We then called for participants for our study by sharing the blog posts across programming forums, such as \NJ, JavaScript, cryptography forums, and also with the industrial colleagues who had experience with \NJ runtime environment. 
Eleven people showed their interest in our work and, in the end, eight agreed to voluntary participate in the study.
All the participants were from industry. 
Specifically, four had 2-5, three had 5-10 and one had more than 10 years of experience in programming.
All of the participants had experience in working with \NJ and developed at least one server-side system in this environment.
Half of the participants expressed that they are ``partly knowledgeable'' in cryptography, meaning that they had an indefinite opinion about some fields of cryptography. 
The remaining participants stated that they are knowledgeable in this field, implying that they are familiar with cryptographic APIs and how and where to apply them. 
None of the participants expressed that they are ``very knowledgeable'' or ``not knowledgeable'' in cryptography. 
Three participants had used the \NJ crypto API prior to attending this study.

\subsection{Research questions}

We sought to answer the following research questions:

\begin{itemize}

\item
\textbf{RQ$_1$}: Does adopting \FC increase developer efficiency when working on a cryptographic task?

\item
\textbf{RQ$_2$}: Is \FC effective in developing secure cryptographic solutions?

\end{itemize}

The purpose of RQ$_1$ was to understand whether developers can complete a cryptographic task faster when they use \FC versus when they rely on the native Crypto API in \NJ.
In RQ$_2$, we looked at developer solutions through the lens of security to identify whether developers make fewer mistakes when they use \FC versus when they use the native \NJ Crypto API.


\subsection{Study procedure}

Prior to the study, all participants were given an introduction to the basic concepts pertinent to the three cryptographic tasks. Particularly, we made sure that they understand both the meaning and the purpose of ``hashing'' and ``encryption'', and they appreciate the difference between symmetric and asymmetric cryptography.
We then asked the participants to fill out a pre-survey to tell us about their years of experience in programming, level of knowledge in cryptography, previous experience with \NJ crypto API, and finally how comfortable they are with the introduced cryptographic concepts.

We attempted to minimize the impact of wrong assumptions with regard to how subjects report their level of knowledge in cryptography by providing them with an explanation of each level with regard to APIs such as \texttt{Hash}, \texttt{Cipher}, \texttt{Sign}, and concepts behind Secret Key, IV and Salt.

\newpage

\begin{itemize}

\item Very knowledgeable: ``I am proficient in cryptography; I fully understand these concepts and am able to use these APIs correctly''.

\item Knowledgeable: ``I am familiar with cryptography fundamentals; I know the purpose of these APIs and fairly understand these concepts, however, I am not confident to develop a secure solution on my own''.

\item Partly knowledgeable: ``I have vague ideas about cryptography; I know the basics \eg what these APIs are, but I rarely developed a cryptographic solution''.

\item Not knowledgeable: ``I am not familiar with cryptography and have not dealt with a cryptographic task''.

\end{itemize}

Each task comprised two activities: (i) solving the task with the \NJ Crypto API; and (ii) providing a solution using \FC.
The participants could consult the official \NJ documentation or other online information sources such as the \SO website, but in case of \FC the only available reference was the project documentation on \GH.

Participants had to start from the first task and fill out a post-survey before moving to the next task.
We did not specify a deadline for the completion of a task, and the subjects were asked to fill out the survey regardless of whether or not they could complete a task.
We collected the following information via a Google form for each activity related to a task:

\begin{itemize}
	
	\item Time taken to work on an activity.
	\item Perceived level of difficulty expressed as a ten-point Likert scale
with 1 being the lowest value and 10 being the highest value.
	\item Feedback esp. regarding problems encountered while working on an activity.
	\item Code piece developed by the subject.
		
\end{itemize}

In addition to satisfaction which is a subjective feeling, 
we asked the participants to consider a few factors when rating the level of difficulty to perform a task.
Specifically, 
how easily you learned about the API;
how easily you could map a task into API elements; 
how obvious the purpose of the APIs were from their names;
and the level to which you were able to find errors in code and debug it.

Finally, subjects were debriefed on the goal of this study and were asked to sign their consent if they agree that we use their anonymised data to report in a publication.

\subsection{Statistical testing}

We ran Wilcoxon signed-rank tests to investigate whether there is a significant difference between the participants' performance when they use \NJ cryptography API versus \FC.
The Wilcoxon signed-rank test is a nonparametric statistical hypothesis test for dependent samples and is used when distributions are not assumed to be normal. 

For each test, we formulated a null hypothesis to determine whether there
is a significant difference between the two activities in each task, \ie when using the \FC versus \NJ cryptography API.
If, after testing a null hypothesis, we find that we can reject the null
hypothesis with a high confidence (\ie $\alpha$= 0.05), we accept
its alternative hypothesis.
%

An example null hypothesis is H$_0$: ``\emph{there is no significant difference in the efficiency of participants when they rely on \NJ crypto API versus \FC for developing cryptographic solutions}''; and the corresponding alternative hypothesis is H$_1$: ``\emph{The efficiency of participants significantly increases when they rely on \FC versus \NJ crypto API for developing cryptographic solutions}''.
The remaining null and alternative hypotheses are analogous.

\section{Study Result}
\label{sec:validationResult}

\begin{figure*}
\begin{subfigure}[b]{0.5\textwidth}
\begin{tikzpicture}
\begin{axis}
[
width=0.8\linewidth,
ybar,
enlargelimits=0.3,
ylabel={Time (minutes)},
xtick=data,
xticklabels={Task 1, Task 2, Task 3}
]

\addplot[black, pattern color=red!75!white, pattern=crosshatch] coordinates {
    (1,6.50) 
    (2,14.72) 
    (3,16.58) 
};

\addplot[black, fill=green!35!gray] coordinates {
    (1,4.87) 
    (2,9.25) 
    (3,7.42) 
};
\end{axis}
\end{tikzpicture}
\caption{Mean time spent on each task}
\label{img:resultTime}
\end{subfigure}
\begin{subfigure}[b]{0.5\textwidth}
\begin{tikzpicture}
\begin{axis}
[
width=0.8\linewidth,
ybar,
enlargelimits=0.3,
legend style={legend columns=-1, draw=none, column sep=1ex},
ylabel={Difficulty (1-10)},
xtick=data,
xticklabels={Task 1, Task 2, Task 3}
]
\addplot[black, pattern color=red!75!white, pattern=crosshatch] coordinates {
    (1,2.75) 
    (2,5.83) 
    (3,8) 
};

\addplot[black, fill=green!35!gray] coordinates {
    (1,2.75) 
    (2,
    5.40)
    (3,4.57) 
};
\legend{\NJ API,\FC}
\end{axis}
\end{tikzpicture}
\caption{Perceived level of difficulty in each task}
\label{img:resultDifficulty}
\end{subfigure}
\caption{The study result}
\label{img:studyresult}
\end{figure*}
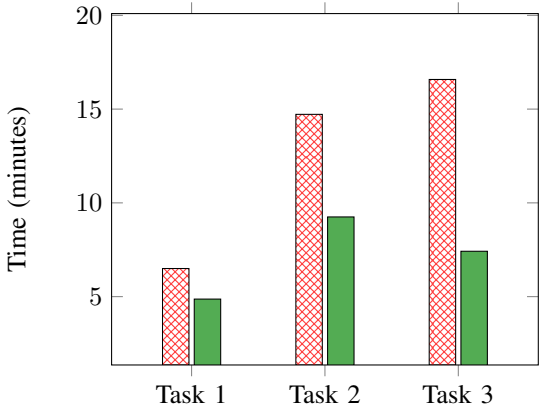
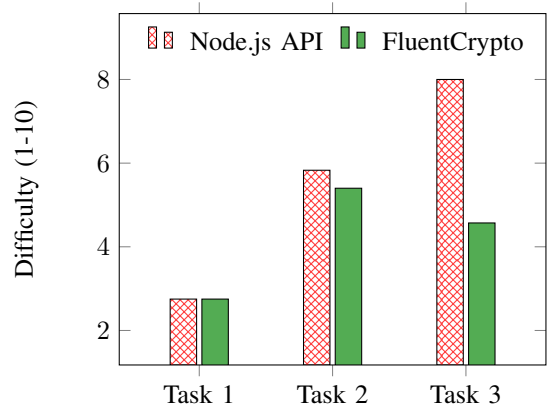

This study showed that \FC helped participants to complete more crypto tasks in a shorter time compared to when they used the \NJ crypto API.
~\autoref{img:studyresult} presents the mean time that participants spent on each task as well as the perceived level of difficulty when dealing with each task. 
Notably, participants spent on average 43\% shorter time to complete each task when they relied on \FC as compared to \NJ cryptography API.
We also found that the participants succeeded to complete five more tasks when they relied on \FC.
The investigation of the solutions revealed that there was no security issue in the ones developed for the first task. 
Only three solutions were secure when participants used \NJ cryptography API to develop the second and third tasks, whereas every solution developed with \FC was secure.

We applied statistical tests to determine whether these findings are significant.  
We found that the p-values of all tests are smaller than 0.05 indicating that we can reject the null hypotheses. 
In other words, there is sufficient evidence with a high confidence to conclude that participants are more efficient to develop secure solutions when adopting \FC as compared to the \NJ crypto API.

In general, participants perceived a higher level of difficulty as the tasks became more complex.
We noted that the subjects mostly relied on the built-in crypto operations in \FC which handle low-level complexities.
This could explain why the gain of using \FC increased with more complex task, especially, for novices who lacked programming experience or had limited crypto knowledge.
Unexpectedly though, when dealing with Task 2, a couple of experienced developers perceived a higher level of difficulty with \FC than when they used the \NJ standard crypto library.
However, these developers often completed their tasks faster when they used \FC.
%
%
Moreover, participants rated the third task, on average, easier and spent shorter time on this task than the second task when they used the \FC.
Although these two tasks are different, \FC is designed in such a way that the solutions are similar to implement.
Therefore, what the subjects experienced earlier, especially when dealing with the second task, may have contributed to their success in the third task.




In the following, we discuss the result of each task based on participant characteristics such as years of experience in programming, level of self-reported knowledge in cryptography, and prior experience with the \NJ crypto API.

\subsection{Does adopting \FC increase developer efficiency when working on a cryptographic task?}

We discuss our findings regarding the time that participants invested in each task as well as 
the perceived level of difficulty.

\subsubsection{Task 1}


\begin{table}[]
\caption{Task 1 result}
\subcaption{Years of programming experience}\label{Task1:1}
\begin{tabular}{| l |c|c|c|c|}
\hline
\multirow{2}{*}{\textbf{Years of Experience}} & \multicolumn{2}{c|}{\textbf{Node.js API}} & \multicolumn{2}{c|}{\textbf{FluentCrypto}} \\ \cline{2-5} 
& \textbf{Time}       & \textbf{Difficulty}      & \textbf{Time}         & \textbf{Difficulty}	\\ \hline
2-5 years & 9:15 min & 3.5 & 5 min & 3.5\\
\hline
5-10 years & 4:20 min & 2 & 4:20 min & 2\\
\hline
$>$ 10 years & 2 min & 2 & 6 min & 2\\
\hline
\end{tabular}

\bigskip
\subcaption{Perceived level of knowledge in cryptography}\label{Task1:2}
\begin{tabular}{| l | c | c | c | c |}
\hline
\multirow{2}{*}{\textbf{Crypto Knowledge}} & \multicolumn{2}{c|}{\textbf{Node.js API}} & \multicolumn{2}{c|}{\textbf{FluentCrypto}} \\ \cline{2-5} 
& \textbf{Time}       & \textbf{Difficulty}      & \textbf{Time}         & \textbf{Difficulty}	\\ \hline
Partly knowledgeable& 9:15 min & 3.5 & 5 min & 3.5\\
\hline
Knowledgeable& 3:45 min & 2 & 4:45 min & 2\\
\hline
\end{tabular}

\bigskip
\subcaption{Experience with \NJ cryptography API}\label{Task1:3}
\begin{tabular}{|l|c|c|c|c|} 
\hline
\multirow{2}{*}{\begin{tabular}[c]{@{}l@{}}\textbf{\NJ}\\ \textbf{Crypto Experience}\end{tabular}} & \multicolumn{2}{c|}{\textbf{Node.js API}} & \multicolumn{2}{c|}{\textbf{FluentCrypto}}  \\ 
\cline{2-5}
                                                                                                      & \textbf{Time} & \textbf{Difficulty}       & \textbf{Time} & \textbf{Difficulty}         \\ 
\hline
No Experience                                                                                         & 8:25 min      & 3                         & 5 min         & 3                           \\ 
\hline
Experienced                                                                                           & 3:20 min      & 2.5                       & 4:40 min      & 2.5                         \\
\hline
\end{tabular}
\end{table}

All participants solved the first task with both the \NJ crypto API and \FC.
\autoref{Task1:1} explains the results regarding task one. The difficulty level looks similar among the groups but the time needed for the task increases for less experienced participants and decreases for more experienced participants.
Nevertheless, the one participant who had more than 10 years' programming experience completed the task with the help of the standard \NJ API three times faster than using the \FC.
\autoref{Task1:2} and \autoref{Task1:3} reveal similar results based on the perceived level of cryptographic knowledge and experience with the \NJ crypto API.
Similarly to our observation from the programming experience perspective, developers who had solid knowledge in cryptography as well as those who were familiar with the \NJ crypto API, completed the task faster than when they used the \FC.
  
\subsubsection{Task 2}

The second task was more challenging for participants than the first task.
The overall perceived difficulty was increased and the task took more time to solve (see \autoref{Task2:2} and \autoref{Task2:3}).
Two participants could not finish the task with \NJ crypto API, whereas everyone completed the task when using \FC.
It is evident that \FC decreased the perceived difficulty for less experienced participants.
The time taken to finalize the task when using \FC reduced for all participant groups except for the group ``5-10 years of experience'' (\autoref{Task2:1}).

\begin{table}
\caption{Task 2 result}
\centering

\subcaption{Years of programming experience}\label{Task2:1}
\begin{tabular}{| l | c | c | c | c |}
\hline
\multirow{2}{*}{\textbf{Years of Experience}} & \multicolumn{2}{c|}{\textbf{Node.js API}} & \multicolumn{2}{c|}{\textbf{FluentCrypto}} \\ \cline{2-5} 
& \textbf{Time}       & \textbf{Difficulty}      & \textbf{Time}         & \textbf{Difficulty}	\\ \hline
2-5 years & 17 min & 6.5 & 10 min & 5.5\\
\hline
5-10 years & 8:20 min & 4 & 11 min & 7\\
\hline
$>$ 10 years & 12 min & 5 & 1 min & 3\\
\hline
\end{tabular}

\bigskip
\subcaption{Perceived level of knowledge in cryptography}\label{Task2:2}
\begin{tabular}{| l | c | c | c | c |}
\hline
\multirow{2}{*}{\textbf{Crypto Knowledge}} & \multicolumn{2}{c|}{\textbf{Node.js API}} & \multicolumn{2}{c|}{\textbf{FluentCrypto}} \\ \cline{2-5} 
& \textbf{Time}       & \textbf{Difficulty}      & \textbf{Time}         & \textbf{Difficulty}	\\ \hline
Partly knowledgeable& 27:20 min & 9 & 11:30 min& 5.5\\
\hline
Knowledgeable & 9:30 min & 4 & 7:40 min& 5.5\\
\hline
\end{tabular}

\bigskip
\subcaption{Experience with \NJ cryptography API}\label{Task2:3}
\begin{tabular}{| l | c | c | c | c |}
\hline
\multirow{2}{*}{\begin{tabular}[c]{@{}l@{}}\textbf{\NJ}\\ \textbf{Crypto Experience}\end{tabular}} & \multicolumn{2}{c|}{\textbf{Node.js API}} & \multicolumn{2}{c|}{\textbf{FluentCrypto}}  \\ 
\cline{2-5}
                                                                                                      & \textbf{Time} & \textbf{Difficulty}       & \textbf{Time} & \textbf{Difficulty}         \\ 
\hline
No Experience & 23 min & 8 & 11 min & 5\\
\hline
Experienced & 9 min & 4.5 & 6:30 min& 6.5\\
\hline
\end{tabular}
\end{table}

\subsubsection{Task 3}

The third task had the highest difficulty for the participants when using the \NJ crypto API (\autoref{Task3:1}).
It also took on average more time to complete by participants than the other two tasks.
\FC decreased the difficulty and time required to complete the task for all groups, especially for the less experienced group (\autoref{Task3:2}).
\FC improved difficulty and the required time to solve the task for both participants with and without knowledge of \NJ crypto API (\autoref{Task3:3}).
One of the senior participants who had beyond 10 years' experience opted out of the study at this phase.
When we asked the reason, he said ``I must deal with an urgent matter at work''.
Of the remaining participants, 
three failed to finish the task with \NJ crypto API, but everyone succeeded in this task when they used \FC.

\begin{table}
\caption{Task 3 result}
\centering

\subcaption{Years of programming experience}\label{Task3:1}
\begin{tabular}{| l | c | c | c | c |}
\hline
\multirow{2}{*}{\textbf{Years of Experience}} & \multicolumn{2}{c|}{\textbf{Node.js API}} & \multicolumn{2}{c|}{\textbf{FluentCrypto}} \\ \cline{2-5} 
& \textbf{Time}       & \textbf{Difficulty}      & \textbf{Time}         & \textbf{Difficulty}	\\ \hline
2-5 years & 20 min & 9 & 9 min & 5\\
\hline
5-10 years & 6:20 min & 5 & 5:20 min& 4\\
\hline
$>$ 10 years & NA  & NA & NA & NA \\
\hline
\end{tabular}

\bigskip
\subcaption{Perceived level of knowledge in cryptography}\label{Task3:2}
\begin{tabular}{| l | c | c | c | c |}
\hline
\multirow{2}{*}{\textbf{Crypto Knowledge}} & \multicolumn{2}{c|}{\textbf{Node.js API}} & \multicolumn{2}{c|}{\textbf{FluentCrypto}} \\ \cline{2-5} 
& \textbf{Time}       & \textbf{Difficulty}      & \textbf{Time}         & \textbf{Difficulty}	\\ \hline
Partly knowledgeable & 20 min & 9 & 9 min & 5\\
\hline
Knowledgeable & 9:45 min& 5.5 & 5:20 min& 4.5\\
\hline
\end{tabular}

\bigskip
\subcaption{Experience with \NJ cryptography API}\label{Task3:3}
\begin{tabular}{| l | c | c | c | c |}
\hline
\multirow{2}{*}{\begin{tabular}[c]{@{}l@{}}\textbf{\NJ}\\ \textbf{Crypto Experience}\end{tabular}} & \multicolumn{2}{c|}{\textbf{Node.js API}} & \multicolumn{2}{c|}{\textbf{FluentCrypto}}  \\ 
\cline{2-5}
                                                                                                      & \textbf{Time} & \textbf{Difficulty}       & \textbf{Time} & \textbf{Difficulty}         \\ 
\hline
No Experience & 17:30 min& 7.5 & 7:25 min & 4.5\\
\hline
Experienced & 9:40 min & 7 & 7:30 min& 5.5\\
\hline
\end{tabular}
\end{table}

\subsection{Is \FC effective in developing secure cryptographic solutions?}

We manually assessed the security of each functioning solution 
against a check list of risks provided by a team of external security experts.
We also checked any hindrances raised by the participants.

\subsubsection{Task 1}

Participants were satisfied by the current \NJ documentation, but had difficulties in finding the right part of documentation for \FC.
All participants chose the \texttt{SHA256} algorithm when they used \NJ cryptography API.
In case of \FC, participants either relied on the default secure algorithm specified by CryRule, or they provided their own choices that were acceptable by \FC.

\subsubsection{Task 2}

The participants mostly complained about the \NJ documentation and mentioned that it is either incomplete or confusing, and consequently, they had to do additional research on the internet. 
When using \FC, 
one participant attempted to generate an IV for an unsupported cipher algorithm and was not satisfied with the thrown error message.


A majority of participants who completed this task misused \NJ crypto API.
Only two participants used a key that was not hardcoded, and others used a hardcoded secret from which they generated the key. 
None of the participants hardcoded a secret when they used \FC.
They mostly relied on the provided call to generate a key.

\subsubsection{Task 3}

In \NJ crypto API, participants were unhappy about a lack of useful examples in the documentation and therefore, had to explore other information sources.
They mainly had difficulties in grasping how to generate a key pair or how to use the keys.
Conversely, developers appreciated the level of details in \FC documentation.
What's more, they found the solution very similar to what they followed while dealing with the previous task.
Three participants also acknowledged that \FC is expressive \eg ``It is great that the name of an API speaks about its functionality'' -- a participant mentioned.


Except one participant, the rest used hard-coded values to generate the key pairs when they used the \NJ cryptography API, which brings severe security risks.
Two solutions also suffered from a small key size.
In contrary, none of the \FC-based solutions suffered from these issues. 
The participants unanimously employed the secure default key-pair generation provided by \FC.


\section{Threats to Validity}
\label{sec:threats}


We had not the possibility to conduct an on-site experiment due to the COVID-19 pandemic.
The participants themselves had to record the time spent on each task. 
It is possible that how each participant calculated the time varies from the actual time.
We mitigated this issue by explaining how to measure the time.
Each participant had to read a task, comprehend basic concepts, and then start the task.

The experience of participants and their familiarity with cryptography could impact the result.
We tried to mitigate this threat by recruiting participants with different level of experience and expertise in this domain, and by briefing them on the basics of cryptography.

Participants self-reported their cryptography knowledge and rated the difficulty of each task based on their personal perceptions of their experience. 
It is possible that each subject had a different perception about his/her experience, which may not have been necessarily true.
Likewise, a perceived ``level of knowledge'' in cryptography may not precisely reflect how developers perform in practice.
We mitigated this threat by defining what each level means.

\newpage

Every participant used both tools, \ie \NJ API and \FC, for each task.
We did not specify which tool to use first, but the one used second may be subject to learning bias.
Therefore, a study with two groups of participants that exclusively use either of these tools may mitigate this threat.
Nevertheless, this has no effect on the security of a provided solution.

We assessed the security of each solution manually, which may be prone to observers' expectancy and subjectivity.
We mitigated this threat by reviewing each solution based on a check list of risks provided by a team of external security experts.

There is a threat to external validity due to the small number of subjects in this study.
We reduced this threat by only recruiting participants who were employed at the software industry.
Nevertheless, future studies with a more representative number of participants is necessary.
What's more, we cannot rule out the possibility that the results could have been different if the tasks were more complex and within a real-world software development context.


\section{Conclusion}
\label{sec:conclusion}

We developed \FC to relieve mainstream developers from dealing with low-level cryptography complexities.
It is built on top of the standard \NJ API and provides a task-based solution \ie developers only state ``what'' they need rather than being concerned about ``how'' to implement a cryptography task.
We also developed a domain-specific language, called CryRule, that crypto experts can use to specify constraints on crypto objects.
\FC relies on these constraints to determine a secure configuration of the API. 
Through an initial study, we found that \FC greatly helps developers to deliver secure solutions in a shorter time.
It prevents common errors by novice developers, but
at the same time it still allows experienced developers to access advanced settings.


This work is the first step toward supporting mainstream developers with a crypto API that is secure by design, and further studies are essential to claim any generalization.
We plan to conduct an extensive experiment with more subjects when COVID-19 restrictions relax and developers come back to their work offices.
Particularly, we are interested to assess the usability of \FC in depth as well as its run-time overhead.
In terms of extending \FC, 
we will investigate whether its current design would support other cryptography tasks as well.
Besides, in this work we relied on a dynamic approach to enforce crypto constraints mainly due to the use of JavaScript, but it is worthwhile to investigate the possibility of adopting statical techniques to provide developers with JIT feedback especially in other settings.





\bibliographystyle{IEEEtran}
\bibliography{fluent}

\end{document}

%% file: preamble.tex
\usepackage{xspace}
\usepackage{graphicx}
\usepackage{amsmath,amssymb}
\usepackage{ifthen}
\usepackage[normalem]{ulem} 
\usepackage{xcolor,colortbl}
\usepackage{hyperref}

\newboolean{showedits}
\setboolean{showedits}{true} 
\ifthenelse{\boolean{showedits}}
{
	\newcommand{\del}[1]{\textcolor{red}{\sout{#1}}} 
	\newcommand{\nbe}[3]{
		{\colorbox{#3}{\bfseries\sffamily\scriptsize\textcolor{white}{#1}}}
		{\textcolor{#3}{\sf\small$\blacktriangleright$\textit{#2}$\blacktriangleleft$}}}
}{
	\newcommand{\del}[1]{} 
	
	\newcommand{\nbe}[3]{}
}

\newboolean{showcomments}
\setboolean{showcomments}{true} 
\newcommand{\id}[1]{$-$Id: scgPaper.tex 32478 2010-04-29 09:11:32Z oscar $-$}

\ifthenelse{\boolean{showcomments}}
 {
 	\newcommand{\nbc}[3]{
 		{\colorbox{#3}{\bfseries\sffamily\scriptsize\textcolor{white}{#1}}}
		{\textcolor{#3}{\sf\small$\blacktriangleright$\textit{#2}$\blacktriangleleft$}}}
	
 }{
 	\newcommand{\nbc}[3]{}
 	
 }

\usepackage[most]{tcolorbox}
\ifthenelse{\boolean{showedits}}
{
  \newtcolorbox{inserted}{%
       title=Inserted text:,
       colframe=blue,colback=blue!5!white,
       breakable,
       leftrule=0mm, 
       bottomrule=0mm,
       rightrule=0mm,
       toprule=0mm,
       arc=0mm, outer arc=0mm,
       oversize
  }
  \newtcolorbox{deleted}{%
       title=Deleted text:,
       colframe=red,colback=red!5!white,
       breakable,
       leftrule=0mm, 
       bottomrule=0mm,
       rightrule=0mm,
       toprule=0mm,
       arc=0mm, outer arc=0mm,
       oversize
  }
  \newtcolorbox{refactored}{%
       title=Rewritten text:,
       colframe=blue,colback=red!5!white,
       breakable,
       leftrule=0mm, 
       bottomrule=0mm,
       rightrule=0mm,
       toprule=0mm,
       arc=0mm, outer arc=0mm,
       oversize
  }
}{

}
\newboolean{isblinded}
\setboolean{isblinded}{true}
\ifthenelse{\boolean{isblinded}}
{\newcommand\blind[1]{BLINDED\xspace}}
{\newcommand\blind[1]{#1\xspace}}

\newcommand{\commented}[1]{}

\newcommand{\eg}{\emph{e.g.,}\xspace}
\newcommand{\ie}{\emph{i.e.,}\xspace}
\newcommand{\etal}{\emph{et al.}\xspace}
\newcommand{\etc}{\emph{etc.}\xspace}

%% file: fluent.bbl
\begin{thebibliography}{10}
\providecommand{\url}[1]{#1}
\csname url@samestyle\endcsname
\providecommand{\newblock}{\relax}
\providecommand{\bibinfo}[2]{#2}
\providecommand{\BIBentrySTDinterwordspacing}{\spaceskip=0pt\relax}
\providecommand{\BIBentryALTinterwordstretchfactor}{4}
\providecommand{\BIBentryALTinterwordspacing}{\spaceskip=\fontdimen2\font plus
\BIBentryALTinterwordstretchfactor\fontdimen3\font minus
  \fontdimen4\font\relax}
\providecommand{\BIBforeignlanguage}[2]{{%
\expandafter\ifx\csname l@#1\endcsname\relax
\typeout{** WARNING: IEEEtran.bst: No hyphenation pattern has been}%
\typeout{** loaded for the language `#1'. Using the pattern for}%
\typeout{** the default language instead.}%
\else
\language=\csname l@#1\endcsname
\fi
#2}}
\providecommand{\BIBdecl}{\relax}
\BIBdecl

\bibitem{ibm}
T.~IBM and the Ponemon~Institute, ``Cost of a data breach report,'' Tech. Rep.,
  2020.

\bibitem{Patnaik2020}
\BIBentryALTinterwordspacing
N.~Patnaik, J.~Hallett, and A.~Rashid, ``Usability smells: An analysis of
  developers{\textquoteright} struggle with crypto libraries,'' in
  \emph{Fifteenth Symposium on Usable Privacy and Security ({SOUPS}
  2019)}.\hskip 1em plus 0.5em minus 0.4em\relax Santa Clara, CA: {USENIX}
  Association, Aug. 2019. [Online]. Available:
  \url{https://www.usenix.org/conference/soups2019/presentation/patnaik}
\BIBentrySTDinterwordspacing

\bibitem{hazhirpasand2019impact}
M.~Hazhirpasand, M.~Ghafari, S.~Kr{\"u}ger, E.~Bodden, and O.~Nierstrasz, ``The
  impact of developer experience in using {Java} cryptography,'' in \emph{2019
  ACM/IEEE International Symposium on Empirical Software Engineering and
  Measurement (ESEM)}.\hskip 1em plus 0.5em minus 0.4em\relax IEEE, 2019, pp.
  1--6.

\bibitem{Hazhirpasand2020}
\BIBentryALTinterwordspacing
M.~Hazhirpasand, M.~Ghafari, and O.~Nierstrasz, ``Java cryptography uses in the
  wild,'' in \emph{Proceedings of the 14th ACM / IEEE International Symposium
  on Empirical Software Engineering and Measurement (ESEM)}, ser. ESEM
  '20.\hskip 1em plus 0.5em minus 0.4em\relax New York, NY, USA: Association
  for Computing Machinery, 2020. [Online]. Available:
  \url{https://doi.org/10.1145/3382494.3422166}
\BIBentrySTDinterwordspacing

\bibitem{zhang2018code}
T.~Zhang, G.~Upadhyaya, A.~Reinhardt, H.~Rajan, and M.~Kim, ``Are code examples
  on an online q\&a forum reliable?: a study of api misuse on stack overflow,''
  in \emph{2018 IEEE/ACM 40th International Conference on Software Engineering
  (ICSE)}.\hskip 1em plus 0.5em minus 0.4em\relax IEEE, 2018, pp. 886--896.

\bibitem{Ghafari2017}
M.~{Ghafari}, P.~{Gadient}, and O.~{Nierstrasz}, ``Security smells in
  android,'' in \emph{2017 IEEE 17th International Working Conference on Source
  Code Analysis and Manipulation (SCAM)}, 2017, pp. 121--130.

\bibitem{Gadient2019}
P.~Gadient, M.~Ghafari, P.~Frischknecht, and O.~Nierstrasz, ``Security code
  smells in android icc,'' \emph{Empirical Software Engineering}, vol.~24,
  2019.

\bibitem{kruger2017cognicrypt}
S.~Kr{\"u}ger, S.~Nadi, M.~Reif, K.~Ali, M.~Mezini, E.~Bodden, F.~G{\"o}pfert,
  F.~G{\"u}nther, C.~Weinert, D.~Demmler \emph{et~al.}, ``Cognicrypt:
  supporting developers in using cryptography,'' in \emph{2017 32nd IEEE/ACM
  International Conference on Automated Software Engineering (ASE)}.\hskip 1em
  plus 0.5em minus 0.4em\relax IEEE, 2017, pp. 931--936.

\bibitem{Rahaman2019}
\BIBentryALTinterwordspacing
S.~Rahaman, Y.~Xiao, S.~Afrose, F.~Shaon, K.~Tian, M.~Frantz, M.~Kantarcioglu,
  and D.~D. Yao, ``Cryptoguard: High precision detection of cryptographic
  vulnerabilities in massive-sized java projects,'' in \emph{Proceedings of the
  2019 ACM SIGSAC Conference on Computer and Communications Security}, ser. CCS
  '19.\hskip 1em plus 0.5em minus 0.4em\relax New York, NY, USA: Association
  for Computing Machinery, 2019, p. 2455–2472. [Online]. Available:
  \url{https://doi.org/10.1145/3319535.3345659}
\BIBentrySTDinterwordspacing

\bibitem{Tymchuk2018}
Y.~{Tymchuk}, M.~{Ghafari}, and O.~{Nierstrasz}, ``Jit feedback - what
  experienced developers like about static analysis,'' in \emph{2018 IEEE/ACM
  26th International Conference on Program Comprehension (ICPC)}, 2018, pp.
  64--6409.

\bibitem{Smith2020}
\BIBentryALTinterwordspacing
J.~Smith, L.~N.~Q. Do, and E.~Murphy-Hill, ``Why can{\textquoteright}t johnny
  fix vulnerabilities: A usability evaluation of static analysis tools for
  security,'' in \emph{Sixteenth Symposium on Usable Privacy and Security
  ({SOUPS} 2020)}.\hskip 1em plus 0.5em minus 0.4em\relax {USENIX} Association,
  Aug. 2020, pp. 221--238. [Online]. Available:
  \url{https://www.usenix.org/conference/soups2020/presentation/smith}
\BIBentrySTDinterwordspacing

\bibitem{Corrodi2018}
C.~Corrodi, T.~Spring, M.~Ghafari, and O.~Nierstrasz, ``Idea: Benchmarking
  android data leak detection tools,'' in \emph{Engineering Secure Software and
  Systems}, M.~Payer, A.~Rashid, and J.~M. Such, Eds.\hskip 1em plus 0.5em
  minus 0.4em\relax Cham: Springer International Publishing, 2018, pp.
  116--123.

\bibitem{Ranganath2020}
V.-P. Ranganath and J.~Mitra, ``Are free android app security analysis tools
  effective in detecting known vulnerabilities?'' \emph{Empirical Software
  Engineering}, vol.~25, 2020.

\bibitem{hazhirpasand2020cryptoexplorer}
M.~Hazhirpasand, M.~Ghafari, and O.~Nierstrasz, ``Cryptoexplorer: An
  interactive web platform supporting secure use of cryptography {APIs},''
  \emph{arXiv preprint arXiv:2001.00773}, 2020.

\bibitem{nadi2016jumping}
S.~Nadi, S.~Kr{\"u}ger, M.~Mezini, and E.~Bodden, ``Jumping through hoops: Why
  do {Java} developers struggle with cryptography {APIs}?'' in
  \emph{Proceedings of the 38th International Conference on Software
  Engineering}, 2016, pp. 935--946.

\bibitem{green2016developers}
M.~Green and M.~Smith, ``Developers are not the enemy!: The need for usable
  security {APIs},'' \emph{IEEE Security \& Privacy}, vol.~14, no.~5, pp.
  40--46, 2016.

\bibitem{das2014iv}
S.~Das, V.~Gopal, K.~King, and A.~Venkatraman, ``{IV}= 0 security:
  Cryptographic misuse of libraries,'' \emph{Massachusetts Institute of
  Technology, Final Rep}, vol.~6, 2014.

\bibitem{Kruger2020}
\BIBentryALTinterwordspacing
S.~Kr\"{u}ger, K.~Ali, and E.~Bodden, ``Cognicrypt$_{GEN}$: Generating code for
  the secure usage of crypto apis,'' in \emph{Proceedings of the 18th ACM/IEEE
  International Symposium on Code Generation and Optimization}, ser. CGO
  2020.\hskip 1em plus 0.5em minus 0.4em\relax New York, NY, USA: Association
  for Computing Machinery, 2020, p. 185–198. [Online]. Available:
  \url{https://doi.org/10.1145/3368826.3377905}
\BIBentrySTDinterwordspacing

\bibitem{Nguyen:2017}
\BIBentryALTinterwordspacing
D.~C. Nguyen, D.~Wermke, Y.~Acar, M.~Backes, C.~Weir, and S.~Fahl, ``A stitch
  in time: Supporting android developers in writingsecure code,'' in
  \emph{Proceedings of the 2017 ACM SIGSAC Conference on Computer and
  Communications Security}, ser. CCS '17.\hskip 1em plus 0.5em minus
  0.4em\relax New York, NY, USA: Association for Computing Machinery, 2017, p.
  1065–1077. [Online]. Available:
  \url{https://doi.org/10.1145/3133956.3133977}
\BIBentrySTDinterwordspacing

\bibitem{Singleton2020}
\BIBentryALTinterwordspacing
L.~Singleton, R.~Zhao, M.~Song, and H.~Siy, ``Cryptotutor: Teaching secure
  coding practices through misuse pattern detection,'' in \emph{Proceedings of
  the 21st Annual Conference on Information Technology Education}, ser. SIGITE
  '20.\hskip 1em plus 0.5em minus 0.4em\relax New York, NY, USA: Association
  for Computing Machinery, 2020, p. 403–408. [Online]. Available:
  \url{https://doi.org/10.1145/3368308.3415419}
\BIBentrySTDinterwordspacing

\bibitem{Piccolboni2021}
L.~Piccolboni, G.~Di~Guglielmo, L.~P.~Carloni, and S.~Sethumadhavan,
  ``Crylogger: Detecting crypto misuses dynamically,'' in \emph{IEEE Symposium
  on Security \& Privacy (SP) 2021}, 2021.

\bibitem{Gorski2020}
\BIBentryALTinterwordspacing
P.~L. Gorski, Y.~Acar, L.~Lo~Iacono, and S.~Fahl, ``Listen to developers! a
  participatory design study on security warnings for cryptographic apis,'' in
  \emph{Proceedings of the 2020 CHI Conference on Human Factors in Computing
  Systems}, ser. CHI '20.\hskip 1em plus 0.5em minus 0.4em\relax New York, NY,
  USA: Association for Computing Machinery, 2020, p. 1–13. [Online].
  Available: \url{https://doi.org/10.1145/3313831.3376142}
\BIBentrySTDinterwordspacing

\bibitem{FAPI}
\BIBentryALTinterwordspacing
M.~Fowler. (2005) Fluentinterface. [Online]. Available:
  \url{https://www.martinfowler.com/bliki/FluentInterface.html}
\BIBentrySTDinterwordspacing

\end{thebibliography}
